\documentclass{PoS}

\usepackage{amsmath,amssymb,euscript,array,cite,mathrsfs,amsfonts}
\usepackage{epsfig}

\newcommand{\SP}[1]{\begin{equation}\begin{split} #1
\end{split}\end{equation}}
\newcommand{\EQ}[1]{\begin{equation}\begin{split} #1
\end{split}\end{equation}}
\newcommand{\Tr}{\operatorname{Tr}}

\def\PP{\mathscr{P}}
\def\kN{{\cal N}}

\title{QCD with chemical potential on $S^1 \times S^3$}

\ShortTitle{QCD at finite chemical potential in a small hyperspherical box}

\author{Simon Hands\\
        Swansea University\\
        E-mail: \email{s.hands@swansea.ac.uk}}
        
\author{Timothy J. Hollowood\\
        Swansea University\\
        E-mail: \email{t.hollowood@swansea.ac.uk}}

\author{\speaker{Joyce C. Myers}\\%
       Swansea University and University of Groningen\\
       E-mail: \email{j.c.myers@rug.nl}}

\abstract{In this proceedings we summarize our calculation of the phase diagram of QCD at non-zero temperature and chemical potential on $S^1 \times S^3$ from one-loop perturbation theory \cite{Hands:2010zp}, which is valid in the limit $R << \Lambda_{QCD}^{-1}$, where $R$ is the radius of $S^3$. We calculate several observables including the Polyakov lines and the quark number, for large number of colors $N$ and large number of quark flavors $N_f$, on $S^1 \times S^3$, and compare with results for the same system with $N = 3$, and with results for $N=2$ lattice QCD. For $N > 2$ the action is complex and the dominant contributions to the path integral occur in the space of complexified gauge field configurations. This results in the expectation values of the eigenvalues of the Polyakov line lying off the unit circle and out in the complex plane. This is an important issue for the lattice, and also for the calculation on $S^1 \times S^3$ in the large $N$ limit where we obtain analytical results using the saddle point approximation. It is thus necessary to adapt available techniques to locate the stationary solutions in the complexified gauge field configuration space.}

\FullConference{The XXVIII International Symposium on Lattice Filed Theory\\
		 June 14-19,2010\\
		 Villasimius, Sardinia Italy}

\begin{document}

\section{Introduction}

Calculation of the phase diagram of QCD as a function of temperature and chemical potential is complicated by what is known as the sign problem: the action of QCD becomes complex in the presence of a non-zero chemical potential. This prevents conventional methods of lattice simulation because the Boltzmann factor can no longer take a probability interpretation. If the chemical potential is not too large several useful alternative methods of calculating the phase diagram from lattice simulations have been developed (for a recent review see \cite{deForcrand:2010ys}). These are valid for $\mu / T \lesssim 1$. Calculation of the phase diagram at asymptotically large chemical potentials and low temperatures is also possible from perturbation theory \cite{Kurkela:2009gj}. The region of the phase diagram that is most difficult to access is that for moderate chemical potentials and low temperatures, where the sign problem is thought to be more severe, and both conventional lattice techniques and conventional perturbation theory are not available (However, simulations using complex Langevin techniques have the potential to probe this region. See \cite{Aarts:2010vk} for a recent report on the progress and issues using this technique in the XY model.).

At the moment, to calculate in the regime of moderate chemical potentials and low temperatures a sacrifice is necessary. We have opted to sacrifice the large volume limit, compactifying the spatial volume such that perturbation theory becomes valid ($R << \Lambda_{QCD}^{-1}$ where $R$ is the radius of the $S^3$). On a hyperspherical manifold, $S^1 \times S^3$ (as opposed to $(S^1)^4$), small volume results for the phase diagrams of related theories, such as Yang-Mills theory and adjoint QCD, qualitatively resemble lattice results (see for example \cite{Aharony:2003sx,Hollowood:2009sy}). A thermodynamic limit of sorts is obtained by taking $N \rightarrow \infty$. In this case sharp, well-defined phase transitions are possible, even in finite spatial volume. Thus it sometimes happens that larger volume lattice results of a small $N$ theory more closely resemble the small volume large $N$ theory, than the small volume theory of the same $N$. It is useful to consider both perturbative results and lattice results to distinguish between small volume effects, lattice artifacts, large $N$ effects, and non-perturbative contributions. This proceedings reviews our perturbative results for QCD on $S^1 \times S^3$ \cite{Hands:2010zp}, comparing with lattice results for $N = 2$ \cite{Hands:2010vw}.

\section{Background}

For our perturbative calculations, all quantities are derived from the 1-loop action of QCD on $S^1 \times S^3$, which was originally derived in \cite{Aharony:2003sx, Sundborg:1999ue} for theories with more general matter content, and summarized for QCD in \cite{Hands:2010zp}. For QCD with $N$ colors and $N_f$ quark flavors with mass $m$ and chemical potential $\mu$ and at a temperature $T = 1 / \beta$ the action is given by
\SP{
S(\theta_i) = &\sum_{n=1}^{\infty} \frac{1}{n} \left( 1 - z_v (n \beta/R) \right)\sum_{i,j=1}^N\cos(n(\theta_i-\theta_j))\\
&+ \sum_{n=1}^{\infty} \frac{(- 1)^n}{n} N_f z_f (n \beta/R, m
R)\sum_{i=1}^N 
\left[ e^{n \beta \mu+in\theta_i} + 
e^{-n \beta \mu-in\theta_i} \right] ,
\label{pop}
}
where the $\theta_i$ are the angles of the Polyakov line matrix $P = {\rm diag} \{ e^{i \theta_1}, ... , e^{i \theta_N} \}$, and $z_v$, $z_f$ are the single particle partition functions for (vector) gauge fields and fermions, respectively, defined by
\EQ{
z_v (\beta/R)=\sum_{\ell=1}^\infty d_\ell^{(v,T)}e^{-\beta\varepsilon_\ell^{(v,T)}} = 2\sum_{\ell=1}^\infty\ell(\ell+2)e^{-\beta(\ell+1)/R}\ ,
\label{jhh}
}
\EQ{
z_f(\beta/R,mR)=\sum_{\ell=1}^\infty d_\ell^{(f)}e^{-\beta\varepsilon_\ell^{(f)}}=2\sum_{\ell=1}^\infty\ell(\ell+1)e^{-\beta \sqrt{(\ell+\frac12)^2+m^2R^2}/R}\ .
}

\section{$N = 3$ results: Quark number ${\mathscr N}$ and Polyakov lines $\PP_1$ and $\PP_{-1}$}

\begin{figure}[t]
  \hfill
  \begin{minipage}[t]{.49\textwidth}
    \begin{center}
\includegraphics[width=0.9\textwidth]{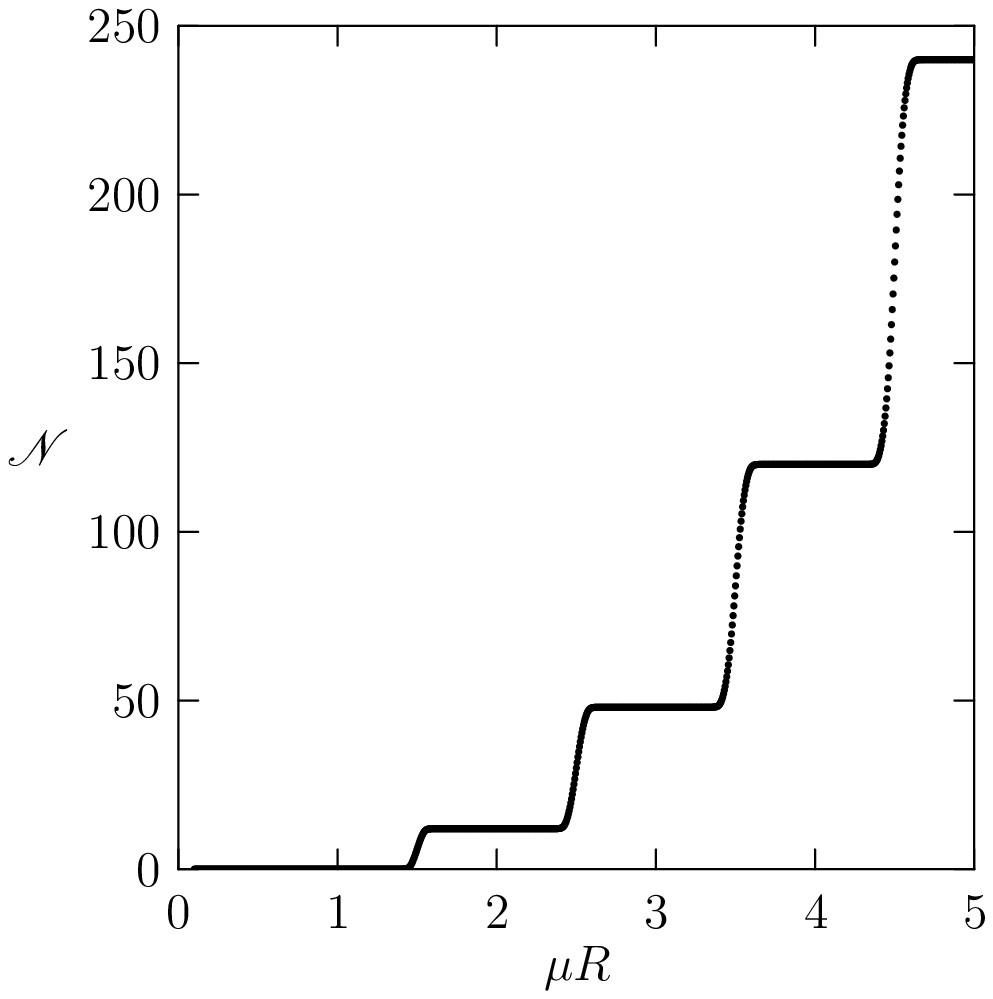}
    \end{center}
  \end{minipage}
  \hfill
  \begin{minipage}[t]{.49\textwidth}
    \begin{center}
\includegraphics[width=0.9\textwidth]{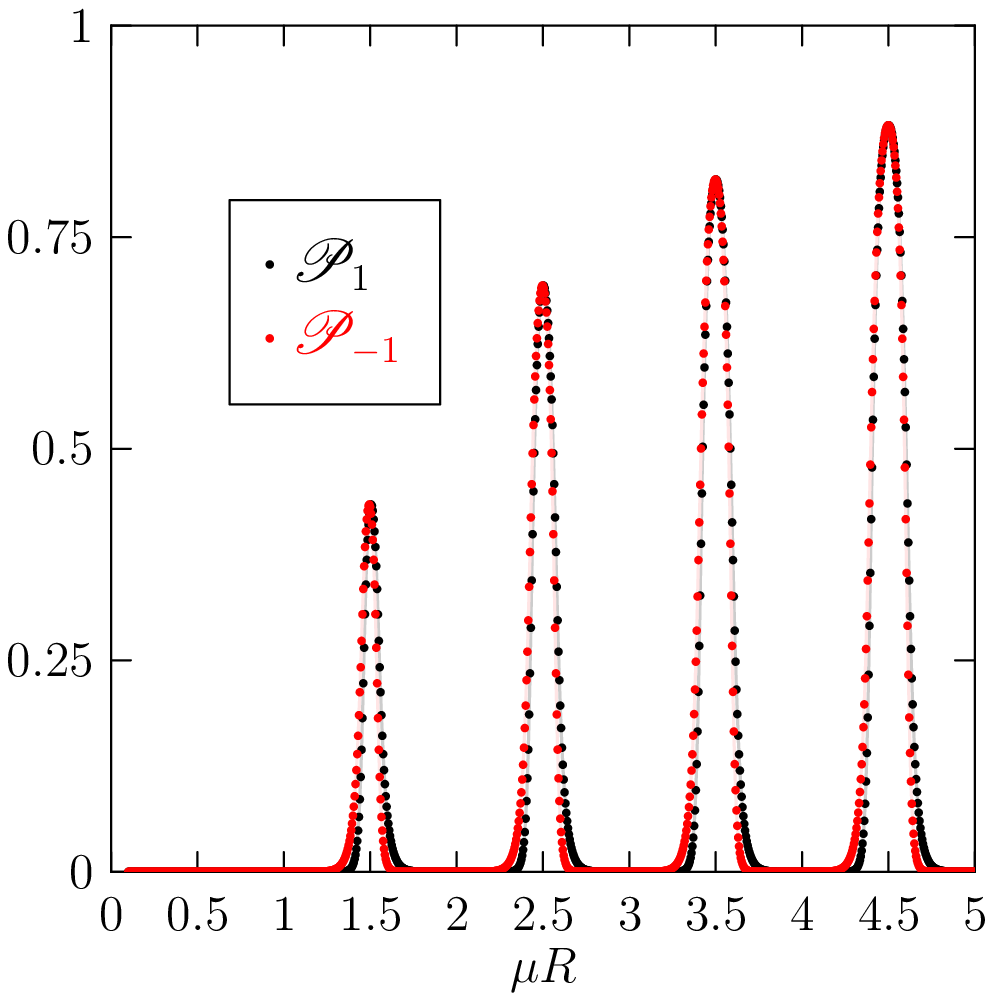}
    \end{center}
  \end{minipage}
  \hfill
\caption{Quark number (Left) and Polyakov lines (Right) as a function of the chemical potential for QCD on $S^1 \times S^3$. (Right). $N = 3$, $N_f = 1$, $m = 0$, $\beta / R = 30$ (low $T$).} 
\label{polys_qnum}
\end{figure}

We are interested in calculating quantities that are derivable from the partition function, which in the low temperature limit takes the form
\EQ{
\begin{aligned}
Z (\beta/R) &= \int \left[ {\mathrm d} \theta \right] \text{exp}\left[ - \sum_{n=1}^{\infty} \frac{1}{n} \left[ \Tr_A (P^n) + (-1)^n N_f z_f (n \beta/R, m R) e^{n \beta \mu} \Tr_F (P^n) \right] \right] ,
\label{partn_fcn}
\end{aligned}
}
where $[d \theta] = \prod_{i=1}^{N} d \theta_i$. We consider a few observables in this proceedings (more are calculated in \cite{Hands:2010zp,Hands:2010vw}), specifically the quark number ${\mathscr N}$ and the Polyakov lines $\PP_1$, $\PP_{-1}$, given by
\begin{eqnarray}
\label{qnum}
&{\mathscr N} &= \frac{1}{\beta} \left( \frac{\partial \log Z}{\partial \mu} \right) \xrightarrow[\beta \rightarrow \infty]{} \frac{N_f}{Z} \int \left[
  {\mathrm d} \theta \right] e^{-S} \sum_{\ell=1}^{\infty}
\sum_{i=1}^{N} 2 \ell (\ell+1) \left[ \frac{e^{\beta \mu}}{e^{\beta
      \mu} + e^{-i \theta_i + \beta \varepsilon^{(f)}_l}} \right],\\
&\PP_{\pm 1} &= \frac{1}{Z} \int \left[ {\mathrm d} \theta \right] e^{-S} \left( \sum_{i=1}^{N} e^{\pm i \theta_i} \right).
\end{eqnarray}
For $N=3$ we are able to calculate these observables numerically because there are only the integrals over $\theta_1$ and $\theta_2$ to compute since $P \in SU(3)$. Figure \ref{polys_qnum} shows the quark number and Polyakov lines as a function of $\mu R$ in the massless limit. The level structure in Figure \ref{polys_qnum} (Left) of the quark number is caused by the Fermi-Dirac distribution function in eq. (\ref{qnum}) which results from taking the $\mu$ derivative. In the low temperature limit ($\beta \rightarrow \infty$) the Fermi-Dirac distribution function is zero for $\mu < \varepsilon^{(f)}_l = (l+\frac{1}{2})/R$, and $1$ for $\mu > \varepsilon^{(f)}_l$, so as $\mu$ passes each energy level the quark number rises another step. Thus, the quark number of the $L$th level is given by ${\mathscr N}_L = N N_f \sum_{\ell=1}^{L} 2 \ell (\ell+1)$ .

Close inspection of Figure \ref{polys_qnum} (Right) shows that the Polyakov lines have the property $\PP_{1} \ne \PP_{-1}^*$. This results at non-zero $\mu$ for $N > 2$ from the fact that the dominant contributions to the path integral lie in the space where the Polyakov line angles $\theta_i$ are complex, a consequence of the complex action. This will be important to our large $N$ analysis since there the saddle point method is valid. It is also interesting that the Polyakov lines as a function of $\mu$ have a deconfinement spike at each energy level $\varepsilon^{(f)}_l$, which results due to the finite separation of the $\varepsilon^{(f)}_l$ on a finite volume and indicates that deconfinement only occurs during quark creation. As $\mu$ is increased past $\varepsilon^{(f)}_l$ the newly created quarks combine and form color singlets.

\section{Large $N$ theory}

Performing the sums over $n$ in eq. (\ref{pop}) we obtain for the action
\EQ{
\begin{aligned}
S(\theta_i)=&\frac12\sum_{i,j=1}^{N} \left[ - \log\sin^2\Big(\frac
{\theta_i-\theta_j}2\Big)+ \sum_{l=1}^{\infty} 2 l (l+2) \log\left[ \cosh (\beta \varepsilon^v_l) - \cos (\theta_i - \theta_j) \right] \right]+N\sum_{i=1}^{N}V(\theta_i)
\end{aligned}
}
where $\varepsilon^v_l = (l+1)/R$ are the energy levels for transverse vectors on $S^1 \times S^3$ and
\EQ{
V(\theta)=i\kN\theta-\sum_{l=1}^{\infty} \sigma_l \left[ \log\left(1+e^{\beta(\mu - \varepsilon^f_l) + i\theta}
\right)+\log\left(1+e^{\beta(- \mu - \varepsilon^f_l) - i\theta}
\right) \right]\ ,
\label{act4}
}
where $\sigma_l \equiv 2 \frac{N_f}{N} l(l+1)$ and we have added the Lagrange multiplier $\kN$ to enforce the
$\det P=1$ constraint, {\it i.e.\/}~$\sum_i\theta_i=0$.

The equation of motion in terms of the Polyakov line eigenvalues $z_i = e^{i \theta_i}$ and the energies $\varepsilon^{(f)}_l$, $\varepsilon^{(v)}_l$, taking $\frac{1}{N} \sum_{i=1}^{N} \xrightarrow[N \rightarrow \infty]{} \int \frac{{\rm d}s}{2 \pi} = \oint_{{\cal C}} \frac{{\rm d}z}{2 \pi i} \varrho(z)$, becomes
\EQ{
\begin{aligned}
&\kN-\sum_{l=1}^{\infty}\sigma_l \left[ \frac{z}{z+e^{-\beta(\mu - \varepsilon^f_l)}} - \frac{e^{-\beta(\mu + \varepsilon^f_l)}}{z+e^{-\beta(\mu + \varepsilon^f_l)}} \right]\\
=&-{\mathfrak P} \int_{{\cal C}} \frac{{\rm d}z'}{2 \pi i} \varrho(z') \left[\frac{z'+z}{z'-z} - \sum_{l=1}^{\infty}\frac{2 l (l+2) \left( z'^2 - z^2 \right)}{(z' - z e^{\beta \varepsilon^b_l}) (z' - z e^{-\beta \varepsilon^b_l})} \right]\ ,
  \end{aligned}
\label{eom2}
}
where ${\cal C}$ is the contour on which the eigenvalues of the Polyakov line lie (following \cite{Gross:1980he,Wadia:1979vk}, but taking the contour to lie off the unit circle). In the confining regions of the $\mu-T$ plane the contour is closed and the most general form of the eigenvalue density $\varrho(z)$, determined from the pole structure of the right hand side of eq. \ref{eom2}, is given by
\EQ{
\varrho(z) = a_{-1} \rho_{-1} + \frac{a_0}{z} + \frac{a_1 \rho_1}{z^2} + \sum_{l=1}^{L} \frac{b_l \rho_{1}}{z+e^{-\beta(\mu - \varepsilon^f_l)}} + \sum_{l=L+1}^{\infty} \frac{b_l \rho_{-1}}{z+e^{-\beta(\mu - \varepsilon^f_l)}} + \sum_{l=1}^{\infty} \frac{c_l \rho_1}{z+e^{-\beta(\mu + \varepsilon^f_l)}} ,
\label{gen_dens_arb_mu}
}
where the factors of $\rho_{\pm1}$ indicate coefficients of terms with or without poles in ${\cal C}$, respectively. Solving the EOM and applying the constraint $\int \frac{{\rm d}z}{2 \pi i} \varrho(z) = 1$  gives the final form
\EQ{
\begin{aligned}
\varrho(z) = & \frac{1}{z} \left[ 1 - \sum_{k=1}^{\infty} (-1)^{k+1} z^{-k} e^{- k \beta \mu} \left( \sum_{l=1}^L b_l \rho_1 e^{k \beta \varepsilon^f_l} \right) \right]\\
& + \frac{1}{z} \sum_{k=1}^{\infty} (-1)^{k+1} \left[ \frac{N_f}{N} \left( \frac{z^{k} e^{k \beta \mu} z_f^{(L+1)} (k \beta / R, m R) + z^{-k} e^{-k \beta \mu} z_f (k \beta / R, m R)}{1 - z_b(k \beta / R)} \right) \right],
\end{aligned}
\label{density}
}
where $b_l \rho_1 = - \sigma_l + 2 \sum_{k=1}^{l-2} k (k+2) b_{l-k-1} \rho_1$, $(l \le L)$, and $z_f^{L+1} = 2 \sum_{l=L+1}^{\infty} l (l+1) e^{- \beta \varepsilon^{(f)}_l}$. The contour ${\cal C}$ is given by $z(s)$ and is obtained by inversion of $\varrho(z) {\rm d}z = i {\rm d}s$, which takes the form
\EQ{
\begin{aligned}
e^{i s} = & z \exp \sum_{k=1}^{\infty} \frac{(-1)^{k+1}}{k} \left[ z^{-k} e^{- k \beta \mu} \left( \sum_{l=1}^L b_l \rho_1 e^{k \beta \varepsilon^f_l} \right) \right]\\
& \times \exp \sum_{k=1}^{\infty} \frac{(-1)^{k+1}}{k} \left[ \frac{N_f}{N} \left( \frac{z^{k} e^{k \beta \mu} z_f^{(L+1)} (k \beta / R) - z^{-k} e^{-k \beta \mu} z_f (k \beta / R)}{1 - z_b(k \beta / R)} \right) \right].
\end{aligned}
\label{contour}
}
In the low temperature limit this reproduces the results in \cite{Hands:2010zp}. In the deconfined phase the eigenvalue density $\varrho(z)$ develops a gap as $\mu$ is increased towards an energy level $\varepsilon^{(f)}_l$. To solve the EOM in this regime we can no longer use Cauchy's theorem in a straightforward way since the contour is not closed. Instead, we must define a resolvent
\EQ{
\omega(z)=-\int_{{\cal C}}\frac{dz'}{2\pi i}\varrho(z')
\frac{z+z'}{z-z'}\ ,
\label{res}
}
and take the eigenvalues of the Polyakov line along a square root branch cut \cite{Gross:1980he,Wadia:1979vk}. The EOM is then obtained from the Plemelj formulae in terms of the average of the resolvent over the cut
\EQ{
zV'(z)=-\frac 12\big[\omega(z+\epsilon)+\omega(z-\epsilon)\big]
\ ,~~~~z\in{\cal C}\ ,
\label{eom4}
}
and the density is obtained from the discontinuity of $\omega(z)$ across the cut
\EQ{
z\varrho(z)=\frac 1{2}\big[\omega(z+\epsilon)-\omega(z-\epsilon)\big]\
,~~~~z\in{\cal C}\ .
}
Thus, observables can be computed from the resolvent by peeling the contour off the cut and collecting the surrounding poles using
\EQ{
\int_{{\cal C}}\frac{dz}{2\pi i}\varrho(z)F(z)=
\oint_{\tilde{\cal C}}\frac{dz}{4\pi i z}\omega(z)F(z)\ .
}

It is now possible to calculate observables in both the confined and deconfined regions. The Polyakov lines can be obtained from $\PP_{\pm 1}=\oint_{{\cal C}}\frac{dz}{2\pi i}\varrho(z) z^{\pm 1}$ in the confined regions, and from an expansion of the resolvent $\omega(z) = \mp 1 \mp 2 \sum_{n=1}^{\infty} z^{\mp 1} \PP_{\pm 1}$ in the deconfined regions. The quark number (normalized by $N^2$) is given by the Lagrange multiplier ${\cal N}$ in the large $N$ limit (as shown in \cite{Hands:2010zp}), which can also be calculated in both the confined and deconfined regions. In the confined regions it is found by solving the equation of motion to obtain the density $\varrho(z)$ and applying the normalization condition $\int \frac{{\rm d}z}{2 \pi} \varrho(z) = 1$ to get an equation for ${\cal N}$. In the deconfined regions the EOM is solved for the resolvent $\omega(z)$, and the $\det P = 1$ constraint $\int \frac{{\rm d}z}{4 \pi i z} \omega(z) \log z = 0$ gives an equation for ${\cal N}$. The full details and results for the calculation of both the Polyakov lines and the quark number are presented in \cite{Hands:2010zp} and resemble what is observed for $N = 3$, with the notable difference that the large $N$ results reveal more sharply defined transitions. As for the order of the transitions in the large $N$ theory, the normalized quark number ${\cal N}$ is continuous as a function of the chemical potential. Also $\partial {\cal N} / {\partial \mu}$ is continuous. But the third derivative of the potential, or $\partial^2 {\cal N} / \partial \mu^2$, is discontinuous, indicating that the transitions are third order, of the Gross-Witten-Wadia type \cite{Gross:1980he,Wadia:1979vk}.

\begin{figure}[t]
  \hfill
  \begin{minipage}[t]{.49\textwidth}
    \begin{center}
\includegraphics[width=0.99\textwidth]{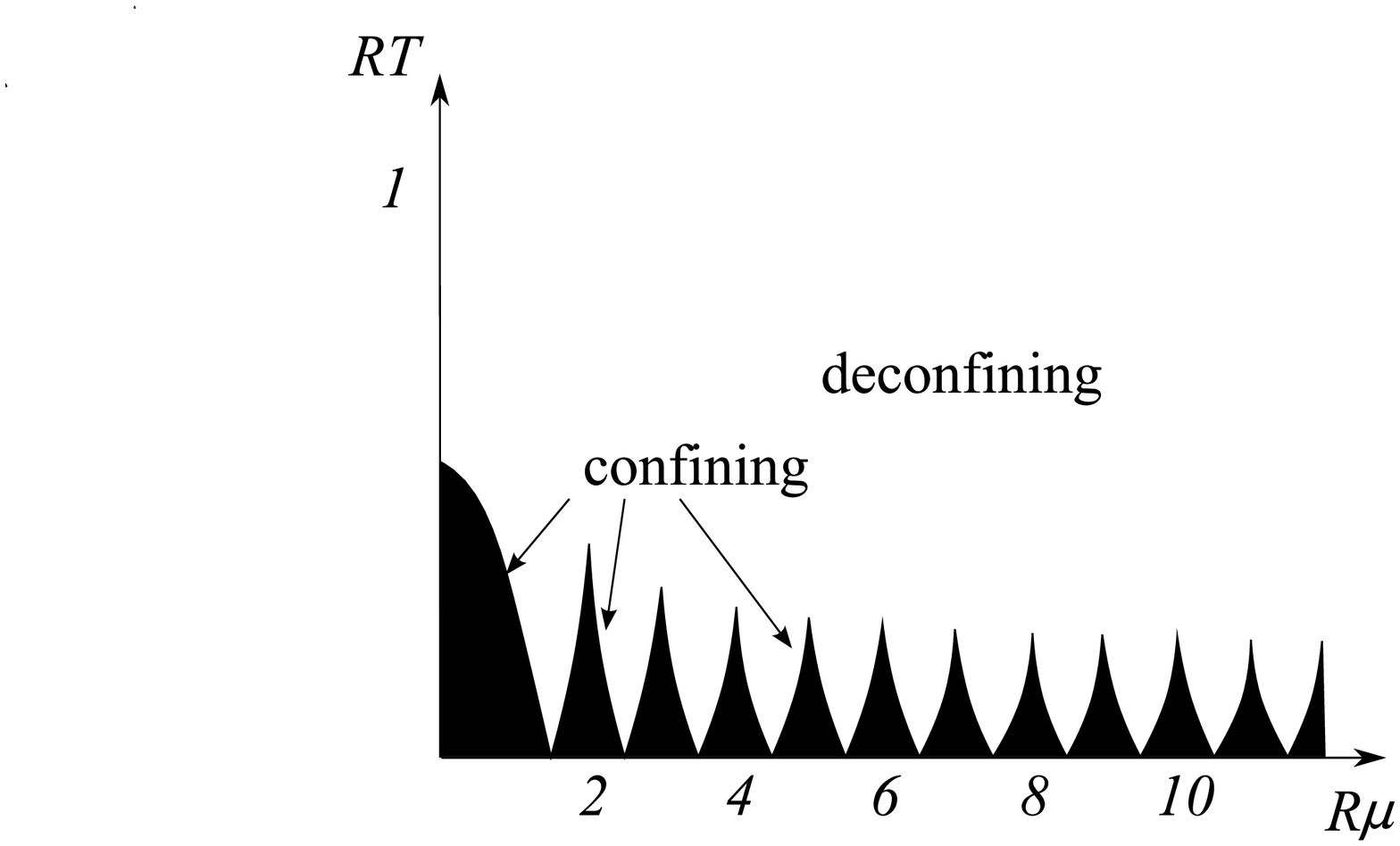}
    \end{center}
  \end{minipage}
  \hfill
  \begin{minipage}[t]{.49\textwidth}
    \begin{center}
\includegraphics[width=0.99\textwidth]{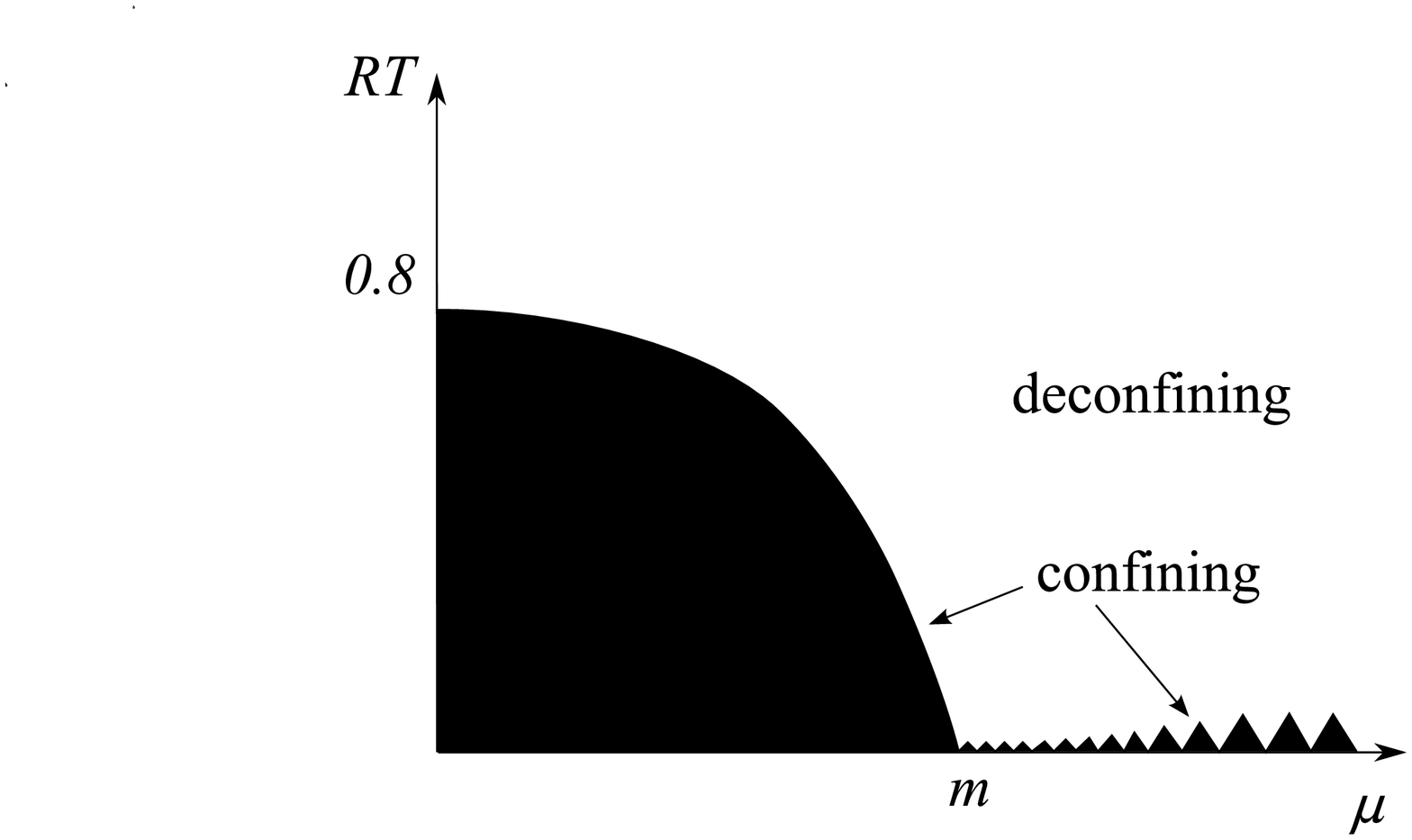}
    \end{center}
  \end{minipage}
  \hfill
\caption{The phase diagram in the $(\mu R, T R)$ plane for QCD at large $N$ and $N_f$ for zero quark mass (Left) and large quark mass (Right).} 
\label{phase_diags}
\end{figure}

The phase diagram in the $\mu R$ - $T R$ plane can be obtained from the equations for the density $\varrho(z)$ (eq. \ref{density}) and the contour $z(s)$ (eq. \ref{contour}) of the Polyakov line eigenvalues in the confining regions. The equations $T(\mu)$ for the lines of transition are obtained by calculating the $z$-values that give $\varrho(z) = 0$, such that a gap is formed in the eigenvalue distribution, and then plugging these into the equation for $z(s)$ such that the gap in the distribution is constrained to lie on the contour ${\cal C}$. The results for small and large quark mass are presented in Figure \ref{phase_diags}, indicating that the series of confinement-deconfinement transitions is a low temperature feature which is delayed as a function of the chemical potential for non-zero quark mass, until that mass is (approximately) reached.

\section{Lattice results}

\begin{figure}[t]
  \hfill
    \begin{center}
\includegraphics[width=0.6\textwidth]{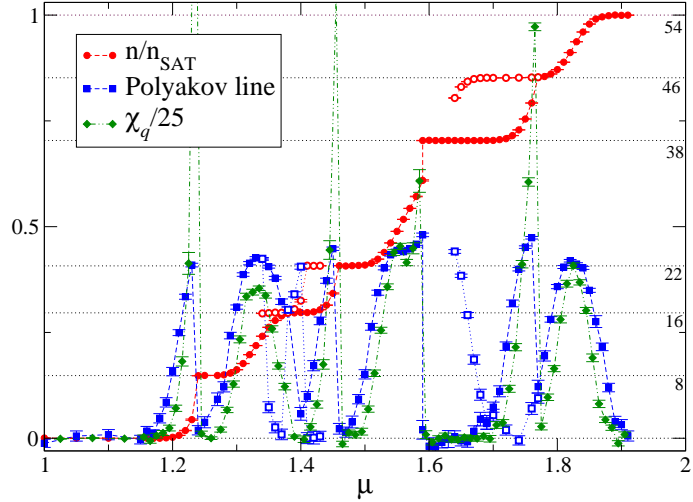}
    \end{center}
\caption{Lattice results from $N = 2$ QCD for the quark number, Polyakov line, and quark number susceptibility as a function of the chemical potential, simulated on a $64 \times 3^3$ lattice with $\beta_g = \frac{2 N}{g^2} = 24$.} 
\label{muscan}
\end{figure}

Lattice results are obtained from simulation of $N = 2$ QCD at non-zero chemical potential and are presented in more detail in \cite{Hands:2010vw}. The main result is reproduced here in Figure \ref{muscan}. These simulations were performed on a $64 \times 3^3$ lattice with a gauge coupling of $\beta_g = \frac{2 N}{g^2} = 24$. The open and closed data points refer to different initial configurations and indicate the existence of multiple stable, or metastable, states. As seen for the theory on $S^1 \times S^3$, the quark number exhibits a level structure and the Polyakov lines show a deconfinement spike at each level transition. It is interesting to note the differences though. From the lattice, the curved shape of the steps in the quark number and matching behaviour in the Polyakov lines could be due to formulating the lattice theory on the torus, or it could be a result of working at stronger coupling. Another interesting feature, which is also observable perturbatively on $S^1 \times S^3$ \cite{Hands:2010vw}, is that the quark number susceptibility $\chi_q \sim T \partial {\cal N} / \partial \mu$ follows the behavior of the Polyakov line, such that it also serves as an indicator of confinement-deconfinement transitions for $\mu \ne 0$.

\section{Relationship between the quark number and Polyakov line}

Finally it is interesting to point out the relationship between the quark number and the Polyakov line which can be obtained in the large mass limit. From eq. (\ref{qnum}), the quark number density goes to
\EQ{
\frac{\mathscr N}{V_3} = 2 N N_f (\frac{m T}{2 \pi})^{3/2} e^{(\mu - m)/T} {\cal P}_1 ,
}
in the large $m$ limit, which agrees precisely with what is obtained in the lattice formulation of $N = 2$ QCD at non-zero density in \cite{Langfeld:1999ig}\footnote{We thank Kurt Langfeld for pointing out this interesting result from the lattice.}. It would be nice to explore this further.

\end{document}